\documentclass[aip,graphicx, groupedaddress, superscriptaddress,longbibliography]{revtex4-1}
\usepackage{graphicx} 
\usepackage{amssymb}
\usepackage{epstopdf} 
\usepackage{color}
\usepackage{hyperref}
\usepackage[utf8]{inputenc}
\usepackage[T1]{fontenc}
\usepackage{mathptmx}

\begin{document}
\title{Contact tribology also affects the slow flow behavior of granular emulsions}


\author{Marcel Workamp}
\author{Joshua A. Dijksman}
\email[]{joshua.dijksman@wur.nl}
\affiliation{Physical Chemistry and Soft Matter, Wageningen University \& Research, Wageningen, The Netherlands}
\date{\today}

\begin{abstract}
Recent work on suspension flows has shown that contact mechanics plays a role in suspension flow dynamics. The contact mechanics between particulate matter in dispersions should depend sensitively on the composition of the dispersed phase: evidently emulsion droplets interact differently with each other than angular sand particles. We therefore ask: what is the role of contact mechanics in dispersed media flow? We focus on slow flows, where contacts are long-lasting and hence contact mechanics effects should be most visible. To answer our question, we synthesize soft hydrogel particles with different friction coefficients. By making the particles soft, we can drive them at finite confining pressure at all driving rates. For particles with a low friction coefficient, we obtain a rheology similar to that of an emulsion, yet with an effective friction much larger than expected from their microscopic contact mechanics. Increasing the friction coefficient of the particles, we find a flow instability in the suspension. Particle level flow and fluctuations are also greatly affected by the microscopic friction coefficient of the suspended particles. The specific rheology of our ``granular emulsions" provides further evidence that a better understanding of microscopic particle interactions is of broad relevance for dispersed media flows.

\end{abstract}
\maketitle
\section{Introduction}\label{sec:intro}
Structured fluids composed of discrete particles, bubbles or droplets are abundant in industry and nature. The importance of these materials is highlighted by the century long continued scientific attention which their flow behavior has received. At sufficient volume fraction of particulate matter, the flow behavior of such structured fluids is generally viewed as consisting of two regimes. In the slow flow limit, interactions are contact-based and the shear stress is rate-independent. At higher driving rates, the material becomes more fluid-like: inertia, collisions or the viscosity of the interstitial fluid~\citep{courrech2003} starts to play a role; the driving stress is then well described by a power law originating mostly from collisional or viscous energy losses. These regimes are often phenomenologically combined by the Herschel-Bulkley (HB) model~\cite{herschel1926}:
\begin{equation}
\displaystyle \tau=\tau_0 + k\dot{\gamma}^n.
\label{floweq:1}
\end{equation}
In this equation, $\tau$ denotes the shear stress, $\tau_0$ the yield stress, $\dot{\gamma}$ the shear rate, $k$ a proportionality constant and $n$ a power law index. The HB model effectively captures the macroscopic flow response of dense granular materials~\cite{jop2006}, emulsions and foams~\cite{paredes2013,dinkgreve2015}, as well as suspensions~\cite{dijksman2010}. Note that for all these systems, the volume fraction $\phi$ has to be high enough in order for the dispersed phase to ``jam'' and resist flow in the slow flow limit~\cite{siemens2010}. The HB constitutive equation also serves as input for flow modeling of amorphous materials deep into the regime where these material seem to be solid-like, in particular as local flow rule in the very successful ``fluidity'' based kinetic elasto-plastic flow modeling \cite{goyon2008,jop2012,kamrin2012}. Even so, although the HB model is applied in a wide variety of materials, exactly how microscopic interactions affect the HB ingredients is an area of active study. There are many microscopic features relevant for the macroscopic flow behavior; the surprising role of roughness, charges, lubrication, adhesion and friction~\cite{johnson2000,zhou2001,lootens2005,becu2006,seto2013, wyart2014, clavaud2017,comtet2017,pons2017,Coulomb2017} have already been suggested especially in faster ``inertial'' flows, where local flow properties can ``turn on'' frictional effects~\cite{seto2013} giving even strong deviations from HB behavior. It is suggested that fluctuations affect $n$ in various regimes~\cite{tighefluct}, but also that microscopic friction coefficients do not significantly affect the HB model~\cite{kamrin2014}.\\ 
\begin{figure}[t!]
\centering
\includegraphics[width=14cm]{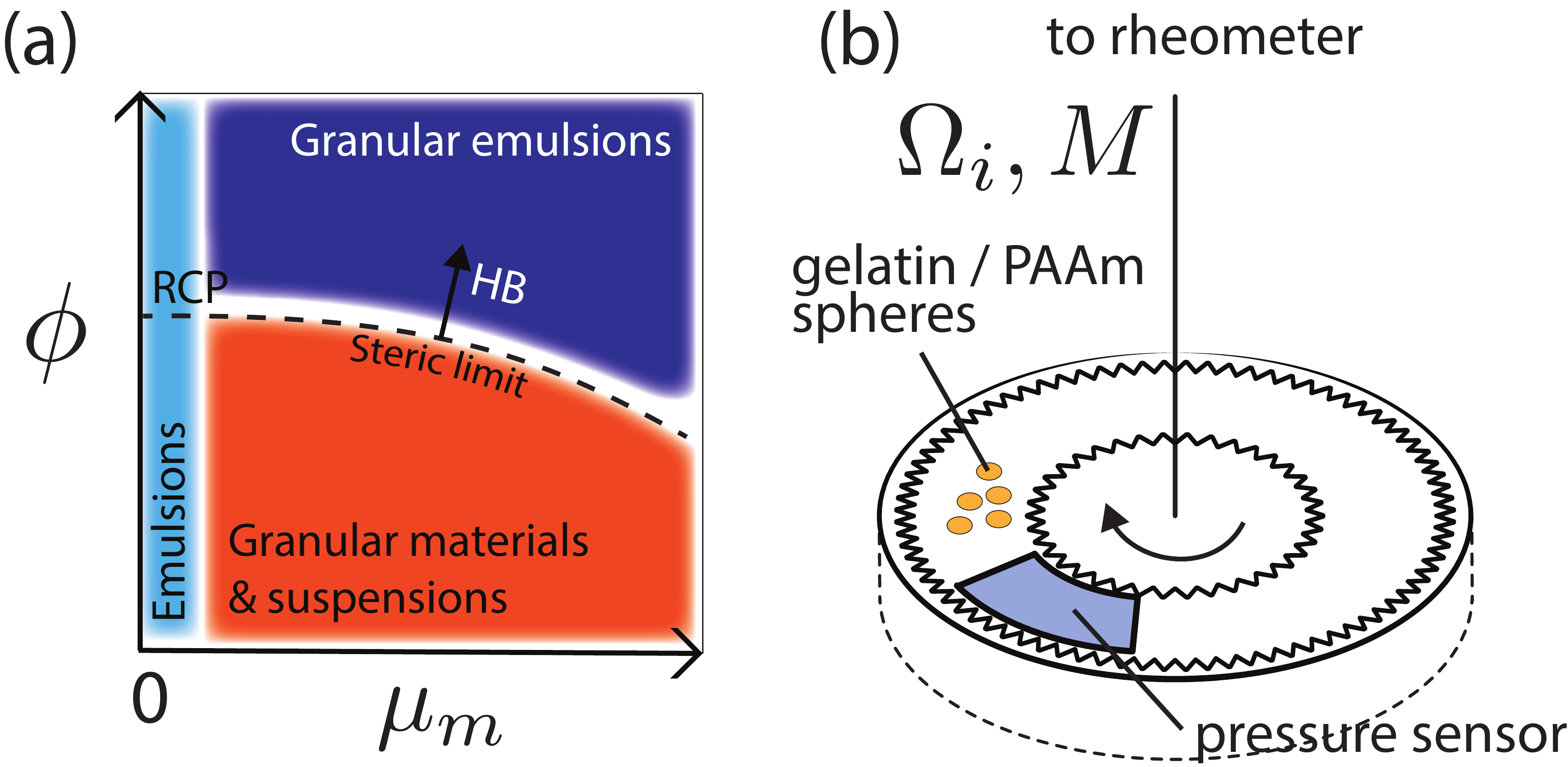}
\caption{\label{friction}(a) Schematic phase diagram. Emulsions are located in the limit of zero microscopic friction constant $\mu_m$. Granular materials and suspensions of solid particles exist at finite $\mu_m$. To obtain HB behavior, the solid fraction $\phi$ of these materials must be above some finite $\phi_{rcp}$ limit, which generally depends on $\mu$~\cite{silbert2010,Hsu2018}, where steric hindrance becomes important. Using soft particles, one can obtain volume fractions above this steric limit for finite $\mu_m$. (b) Schematic of the Couette geometry used in our experiments. $\Omega_i$ is the applied rotation rate, $M$ is the measured torque. Particles are confined to a constant volume environment.}
\end{figure}

Here we show experimentally that microscopic frictional interactions between suspended particles in a dense ``granular emulsion'' have a significant influence even in the \emph{slow} flow limit. We perform experiments using dense suspensions of soft particles confined in fixed volume and sheared in a Couette geometry. Their softness allows us to suspend the particles at high volume fraction ($\phi > \phi_{rcp}$, the random close packing density) while still being able to make them flow. The soft particle suspension we use can therefore be made similarly dense as an emulsion, yet the interactions between the particles are frictional as in a granular material. In our perspective, the granular emulsions we employ exist in the top right corner of the schematic phase diagram sketched in Fig.~\ref{friction}a. Our granular emulsions are therefore rather different from discontinuous shear thickening fluids, as particles are densely packed at all shear rates, at a finite pressure, and therefore always feature semi-permanent contacts among particles.\\ 
\indent We find that granular emulsions have a well defined effective friction coefficient with two peculiar properties: the effective friction coefficient can either be similar, \emph{or} much higher than that of the microscopic coefficient, depending on the magnitude of the microscopic friction coefficient. Furthermore, the effective friction coefficient of the suspension can be rate dependent such that it gets \emph{smaller} at higher shear rates. Even though weak flow instabilities in flowing suspensions have been observed before~\cite{lu2007, dijksman2011}, we find ``yield stress'' reductions of up to a factor two. Our results highlight the importance of understanding the coupling between microscopic interactions and macroscopic flow behavior and their integration in numerical and theoretical modeling approaches for dense particulate media.\\
The paper is set up as follows: in Sec.~\ref{sec:mnm} we first present the flow geometry in which we perform all rheological measurements and a characterization of the custom made hydrogel particles in terms of their size, hardness and frictional properties. In Sec.~\ref{sec:stress} we present an overview of experimental results for shear and confining stress dynamics for various suspension types, at various experimental settings. We make a cross comparison of these results and briefly discuss how pressure controlled experiments compare to our volume controlled experimental data. To gain further insight into the rheology of granular emulsions, we discuss their flow behavior and fluctuations in Sec.~\ref{sec:flow}. An overall discussion and conclusion section follows the presented results.

\section{Materials \& Methods}\label{sec:mnm} 
\subsection{Flow setup}\label{subsec:flowsetup} 
We use a custom, 3D-printed (Stratasys Objet 30) Couette cell to perform the flow experiments, see Fig.~\ref{friction}b and Ref.~\cite{workamp2017}. The inner cylinder has radius $r_i$ = 25~mm, while the outer cylinder has radius $r_o$ = 45~mm, such that the gap $r_o-r_i$ = 20~mm $\approx$ 10$d$ with $d$ the particle diameter. We drive the inner cylinder using a rheometer (Anton-Paar MCR301 or MCR501). Both inner and outer cylinder are made rough with teeth of approximately 2.5~mm to minimize wall slip. The height of the shear cell $L$ = 20~mm~$\approx 10d$; the rheometer measures/provides a torque $M$. There is a top cover on the cell that confines only the particles; the fluid can freely move in and out of the cell. Unless otherwise mentioned, we thus confine the particles to a constant volume in all experiments, while we let the particle pressure adjust to the shear rate and amount of particles added to the volume. We thus measure the pressure exerted by the particles only. We measure the particle pressure $P^p$ on a separate lid embedded in the cover. The lid is attached to a load cell. Solvent can freely flow in and out of the cell through the gap around the pressure-sensing lid and cover, and through the gap between the rotating inner cylinder and the cover. The benefit of this approach is that in typical emulsion rheology experiments, performing constant (particle) volume experiments is impossible due to the size of the droplets involved, and confining stresses are at least partly induced by surface tension at the free boundary, which can assume any shape. Our constant volume experiments allow an effective control and characterization of confining pressure. Experimental protocols and results are discussed in the next section. 

\begin{figure}[t!]
\centering
\includegraphics[width=0.9\linewidth]{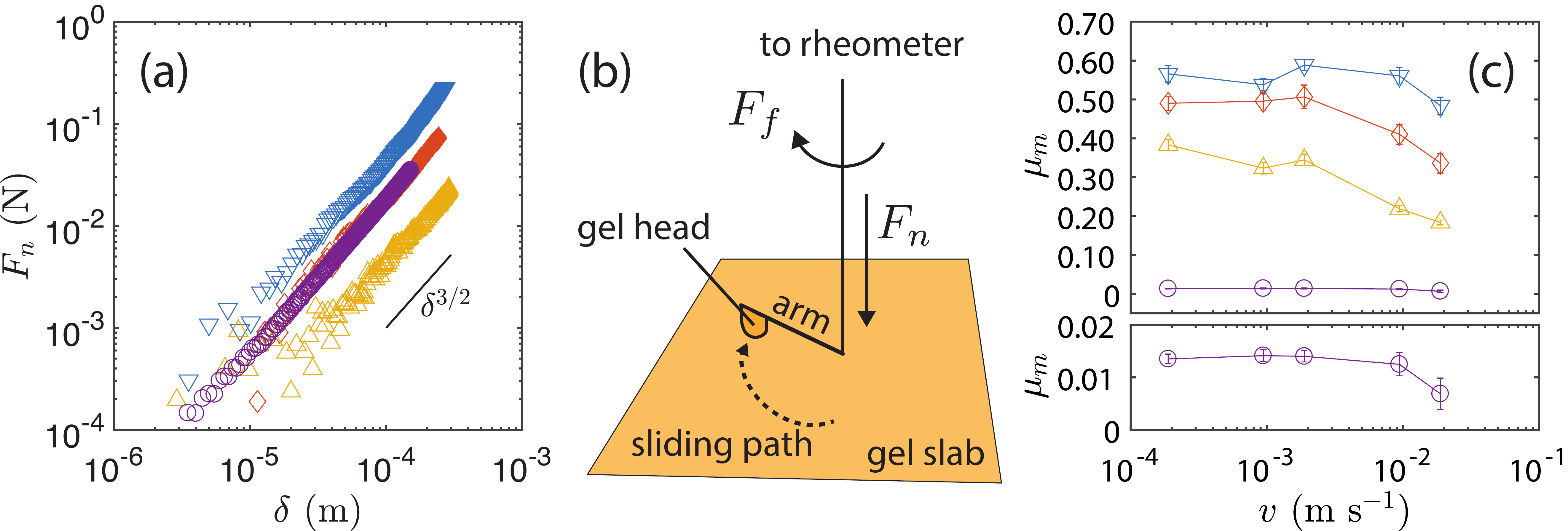}
\caption{\label{fig:MnM} (a) Typical results of uniaxial compression tests of hydrogel particles. Normal force $F_n$ as a function of overlap $\delta$, for a particle of 15\% ($\triangledown$), 10\% ($\diamond$) and 5\% gelatin ($\triangle$), as well as for PAAm ($\circ$). Using Hertzian contact theory, we find the elastic modulus $E$ of the particles as $F_n=\frac{4}{3}\frac{E}{1-\nu^2}R^{1/2}\delta^{3/2}$, where we assume Poisson's ratio $\nu = 0.5$ (incompressible material) and R is the particle radius. (b) Schematic of our tribology setup. (c) Material friction coefficient $\mu_m$ as a function of sliding velocity $v$ for gelatin-gelatin contact (with 15 ($\triangledown$), 10 ($\diamond$), or 5 wt\% gelatin ($\triangle$)) and PAAm-PAAm contacts ($\circ$). Error bars denote the 95\% confidence interval of $\mu_m$, based on linear regression of $F_f(F_n)$.}
\end{figure}

\subsection{Particle hardness}\label{subsec:parthardchar}
We aim to perform experiments on suspensions of macroscopic particles in which we only vary the friction coefficient, while keeping \emph{all} other experimental setting the same. We therefore need to make particles manually, from materials with different surface properties. The materials we choose are hydrogels, because they are soft, can be made through custom synthesis methods~\cite{workamp2016} and are known to have tunable frictional behavior~\cite{gong1998,gong2006,baumberger2006}. In particular, we use low friction polyacrylamide (PAAm) and chemically cross-linked gelatin. We produce bidisperse mixtures with mean diameters $d$ around 2~mm. We make the PAAm particles using a monomer solution that contains 20 wt\% acrylamide and 1 wt\% N,N'-methylenebis(acrylamide) as a cross-linker. We prepare gelatin particles of 5, 10 and 15~wt\% gelatin. To ensure the gelatin particles remain stable to dissolution, we cross-link them with glutaraldehyde~\cite{damink1995glutaraldehyde}. We can keep the composition and the stiffness of the PAAm and gelatin suspensions the same by choosing the right gelatin concentration. To show this, we use uniaxial compression to measure the elastic moduli of the particles. We find the Youngs moduli to be approximately $8.1\times 10^1$ kPa (5\% gelatin), $3.2\times 10^2$ kPa (10\% gelatin), $9.1\times 10^2$ kPa (15\% gelatin) and $3.1\times 10^2$ kPa (PAAm); see Fig.~\ref{fig:MnM}a. Note that the PAAm particles and the 10\% gelatin particles have the same modulus.  

\subsection{Hydrogel friction characterization}\label{subsec:tribo}
We measure the frictional behavior of the hydrogels using a modified version of the ``pin-on-disk'' method. Usually, the pin is held stationary while the disk rotates, see e.g. Ref.~\cite{pitenis2014}. Instead of driving the disk, we drive the pin, a hemispherical gel head (radius 7 mm) using a rheometer (Anton-Paar MCR501). The gel head is securely held on a 3D-printed arm (length $l$ = 3~cm) connected to the rheometer axis, and rubs over a flat gel slab of the same material. The hydrogel samples used for measurement of the friction coefficient have the same chemistry as the particles and are molded using petri-dishes to create flat disks, and silicone rubber (Smooth-On Oomoo) to prepare a hemispherical probe. To ensure a smooth surface of the hemispherical cap, the rubber mold is cast using a ball produced for ball bearing purposes, which have superior smoothness and roundness.\\
\indent We drive the arm at rotation rates ranging from $10^{-3}$ to $10^{-1}$~rps, corresponding to sliding velocities $v$ from $1.9 \times 10^{-4}$ to $1.9 \times 10^{-2}$~m~s$^{-1}$. We measure the torque $M$ and normal force $F_n$ at different heights of the hemispherical probe, to get a range of $F_n$. As the hydrogel surfaces as well as the arm are submersed in water, we correct $F_n$ for buoyancy and $M$ for the viscous contribution of the water. We calculate the frictional force as $F_f = M/l$. We use only the data where 0.02~mN $<F_n<$ 20~mN; in this regime $F_f$($F_n$) is linear and regression yields $\mu_m$. At higher loads, $F_f$ depends more weakly on $F_n$. In our rheology measurements, the particle pressure $P^p$ is around 1~kPa; an estimate of the load on each particle is $P^pd^2$, yielding normal forces in the same range as in our friction measurements. The setup is schematically depicted in Fig.~\ref{fig:MnM}b. Although there is an error associated to measuring a friction coefficient on a circular sliding path rather than in a straight line~\cite{krick2010}, this error is negligible here, since the arm $l$ is much larger than the maximum radius of the contact area ($a_{max}\sim$~1~mm).\\
\indent In Fig.~\ref{fig:MnM}c, we plot $\mu_m$ as a function of the sliding velocity $v$, for the different materials. The errorbars denote the 95\% confidence interval for $\mu_m$. For the polyacrylamide surfaces, the friction coefficient is on the order of 10$^{-2}$ and little effect of $v$ is observed, in agreement with Ref.~\cite{uruena2015}. Only at the highest rate a small decrease can be seen. However, the error bar on this data point is relatively large as the frictional force is small while viscous contributions from the fluid in which the arm is rotating are significant at this rate. Although their results only concern PAAm hydrogels, Ref.~\cite{uruena2015} also helps interpret the polymer concentration dependence of $\mu_m$ for the cross-linked gelatin. The authors show that decreasing the mesh size (i.e. increasing the polymer concentration) of the gel increases its friction coefficient, in agreement with our findings. The friction coefficient of all cross-linked gelatin decreases with $v$. Note that since the modulus and all other particle parameters are the same for PAAm and 10 wt\% gelatin suspensions that we will make, the only difference between them is their frictional behavior.

\section{Stress dynamics}\label{sec:stress}

\subsection{Rate dependence of shear stress and confining pressure} 
To explore the effect of contact friction in suspensions, we measure the shear stress and confining pressure for suspensions made with PAAm and gelatin particles. Composing particles of three different concentrations of gelatin provide us a range of $\mu_m \in \{0.01 \ldots 0.6\}$ as outlined in Sec~\ref{subsec:tribo}. We can confine the suspensions by simply adding more particles in the same volume and measure the resultant confining pressure as a function of shear rate, as is typical for dry granular materials and suspensions~\cite{dacruz2005,boyer2011}. We thus perform measurements at different constant volume fractions. The volume fractions are not known, but we can characterize the density through the pressure $P^p$ at the lowest shear rate $\dot{\gamma}_0 = 1.2 \times 10^{-2}$~s$^{-1}$. Since $P^p$ is finite even at zero shear, we know that $\phi>\phi_{rcp}$, the random close packing density. Due to the size of the particle used, pore fluid flow effects are negligible. It is challenging to match the density of the particles with the solvent as the hydrogel particles are porous to their swelling solvent and their swelling depends on environmental conditions; we therefore use water as the solvent. The maximum hydrostatic pressure can be estimated to be $P_g = \Delta\rho g L \approx$~20~Pa, with $\Delta\rho$ the density difference. As $P^p >> P_g$, we expect no influence of $P_g$. Note that regardless of driving form, $P^p$ in all our experiments on both PAAm and gelatin suspensions is never more than 1.5\% of the modulus: the particles are very weakly compressed and hence remain spherical at all times, so multiple contact effects~\cite{brodu2015,hohler2017} can be neglected. Before measuring the flow curve, we pre-shear the sample at our maximum shear rate ($1.2 \times 10^{2}$~s$^{-1}$) for 10 seconds. After this, we decrease the rate, measuring the required stress for one full rotation of the tool for 41 logarithmically spaced shear rates. Both the top cover and bottom of the cell we use are made of smooth acrylic, so we can perform flow profile measurements via transmission-based particle image velocimetry (see Ref.~\cite{workamp2016} for details). Note that the numerical value used for the shear strese and shear rate in a Couette geometry is subject to some arbitrary choices, due to the inhomogeneity of stress and flow field even at fixed $\Omega_i$; following~\cite{rheochem,chatte2018} we use $\dot{\gamma} \equiv \langle\dot{\gamma}\rangle = \Omega_i\frac{r_o^2+r_i^2}{r_o^2-r_i^2}$ and $\tau \equiv \langle\tau\rangle = M\frac{r_o^2+r_i^2}{4\pi Lr_o^2r_i^2}$; the geometric correction coefficients are all of order one.\\ 

\begin{figure}[t]
\centering
\includegraphics[width=1.0\linewidth]{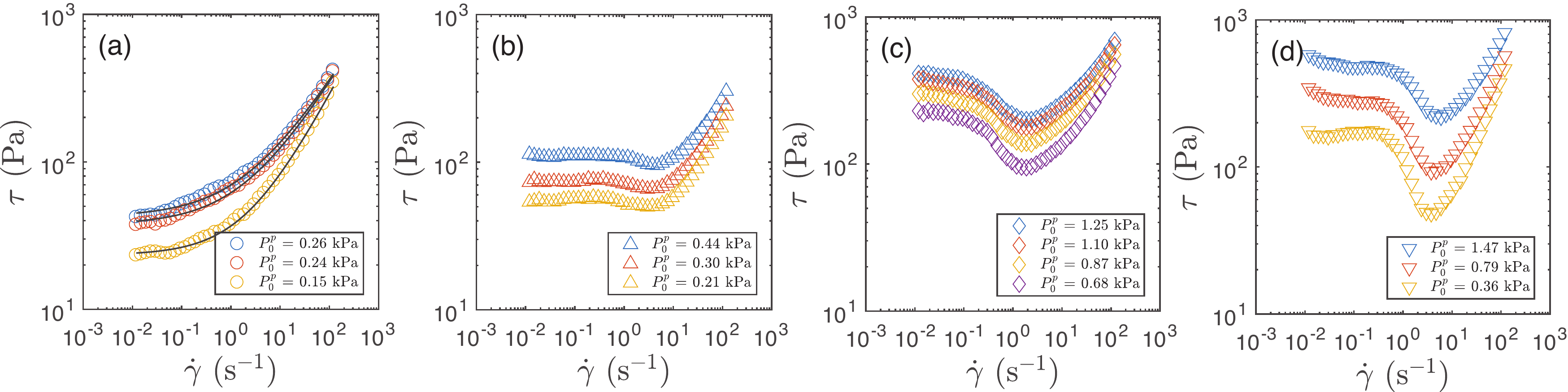}
\caption{\label{fig:flowcurves} Shear stress $\tau$ as a function of shear rate $\dot{\gamma}$ at different volume fraction for PAAm (a), 5\% (b), 10\% (c) and 15\% gelatin (d). We characterize the volume fraction by the pressure $P^p$ at the lowest measured $\dot{\gamma}$. Solid lines represent HB fits according to Eq.~\ref{floweq:1}. From a to d, $\mu_m$ for the particles used increases from 0.01 to 0.6.}
\end{figure}
\begin{figure}[t]
\centering
\includegraphics[width=1.0\linewidth]{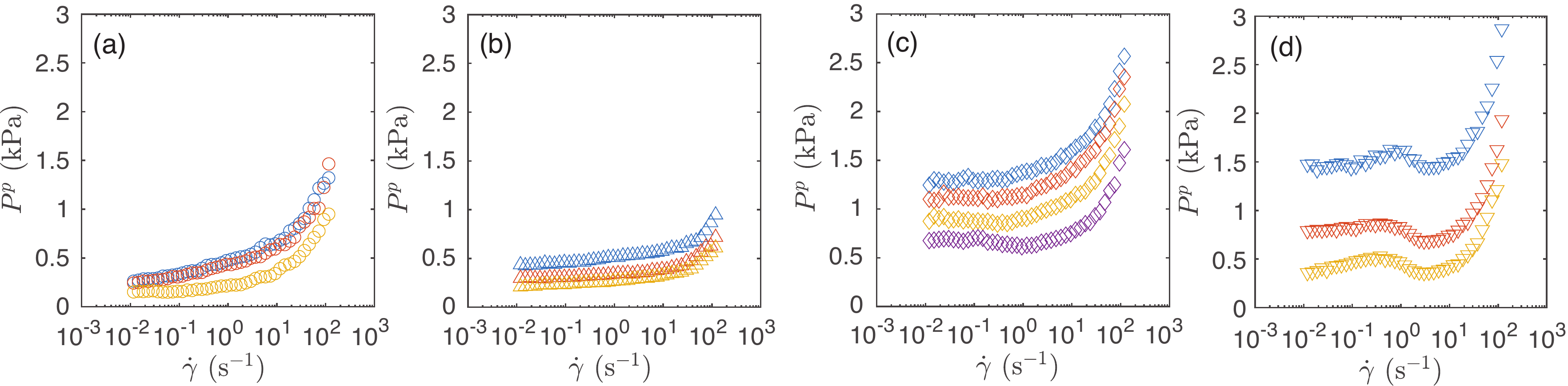}
\caption{\label{fig:Pp_rate} Confining pressure $P^p$ as a function of shear rate $\dot{\gamma}$ at different volume fraction for PAAm (a), 5\% (b), 10\% (c) and 15\% gelatin (d). Same colors/symbols as in Fig.~\ref{fig:flowcurves}.}
\end{figure}

The results of the rate dependent shear stress measurements for all four material types are shown in Fig.~\ref{fig:flowcurves}. Our hydrogel friction measurements indicate that the material friction coefficient $\mu_m$ of PAAm to be approximately 0.01. This means the PAAm particles resemble emulsion droplets: they are deformable and have negligible friction. The rheology of the PAAm suspension is indeed what one may expect for an emulsion: it is well fitted with the HB model (solid lines in Fig.~\ref{fig:flowcurves}a). We find exponents $n$ of about 0.5$-$0.6, similar to what one finds in dense emulsions~\cite{dinkgreve2015}, owing to the deformability of the particles~\cite{barnes1989, brown2014}. The gelatin suspension flow curves are different in character. All gelatin suspension flow curves display non-monotonic behavior, with a distinct minimum or ``dip'' around 5 s$^{-1}$. The dip location seems to be independent of the overall modest pressure variation, but gets more pronounced the higher $\mu_m$ is.\\ 
At the same time during the same experiments, we measure the confining pressure; results are shown in Fig.~\ref{fig:Pp_rate}. Again we find that the PAAm suspension displays a monotonic increase of the confining pressure with the shear rate. For the gelatin suspension, the pressure dynamics is more subtle: at low $\mu_m$ and high pressure, the confining pressure is also monotonically increasing with shear rate. However, as $\mu_m$ increases, a non-monotonicity becomes apparent at low pressure; at the largest $\mu_m$, all pressure dynamics displays this dip. Additionally, at the largest $\mu_m$, the low shear rate dynamics of $P^p$ is weakly rate dependent. Note that the vertical axes in Fig~\ref{fig:Pp_rate} are linear and not logarithmic as those in Fig~\ref{fig:flowcurves}; the dips in $P^p(\dot{\gamma})$ are less pronounced than those in $\tau(\dot{\gamma})$. The pressure measurements are robust; we performed additional experiments in which we measured the confining pressure from the cylinder wall and found the pressure dynamics in the radial direction had the same rate dependence as $P^p$ (not shown).

\subsection{Pressure rescaling}
\begin{figure}[t]
\centering
\includegraphics[width=1.0\linewidth]{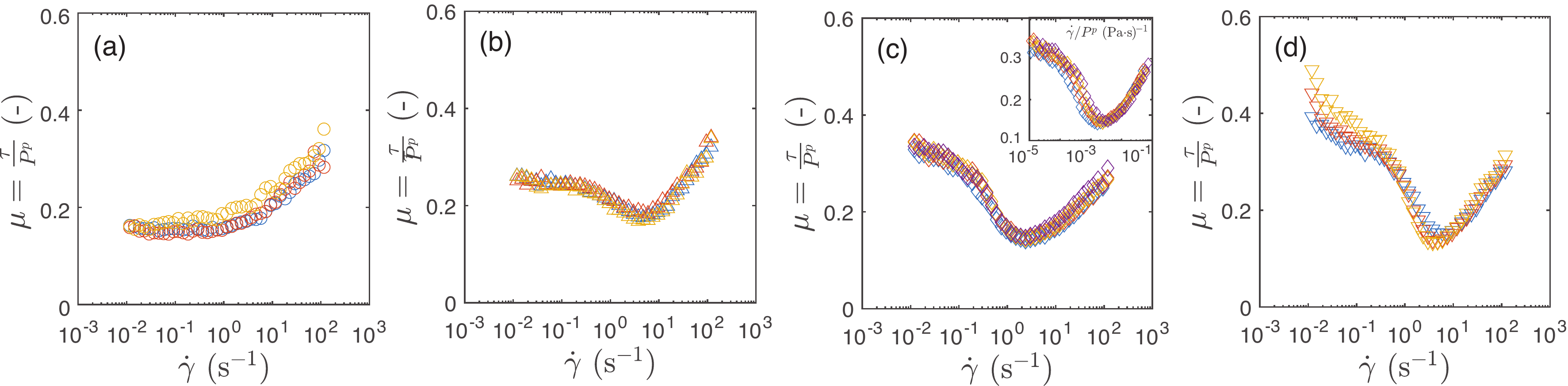}
\caption{\label{fig:MuRate} Effective friction coefficient $\mu$ as a function of shear rate $\dot{\gamma}$ at different pressures for PAAm (a) and gelatin: 5\% (b), 10\% (c) and 15\% (d). Same symbols as in Fig.~\ref{fig:flowcurves}. The inset in c shows the 10\% data as a function of the shear rate rescaled by the pressure.}
\end{figure}

To interpret the flow curves shown in the previous section, we borrow the ideas from dry granular materials~\cite{dacruz2005} and suspensions~\cite{boyer2011}: we investigate the rheology by computing the  \emph{macroscopic} friction coefficient $\mu=\tau(\dot{\gamma})/P^p(\dot{\gamma})$. We combine the shear stress data and the measured confining pressure $P^p$ for all points in the flow curve. We plot $\mu(\dot{\gamma})$ in Fig.~\ref{fig:MuRate}. For all suspensions, the shear stress scales with the confining pressure exerted on the particles and hence we obtain a good collapse of the data obtained at different $P^p$. It is immediately obvious that the PAAm suspension flow behavior in Fig.~\ref{fig:MuRate}a is different from the gelatin suspensions in Fig.~\ref{fig:MuRate}b-d in several ways. We find that for the PAAm suspension, the quasistatic suspension friction coefficient at low shear rates is constant and approximately 0.16. This value is much higher than the material friction coefficient $\mu_m \sim 0.01$. While it has been observed before that even at $\mu_m = 0$, $\mu > 0$ (see for example Refs~\cite{chialvo2012,trulsson2017,peyneau2008}), we find this result counter-intuitive, as it suggests that contact friction is indeed not the main source of dissipation in PAAm suspensions, despite the pressure rescaling. This observation is perhaps related to how the friction coefficient of a rough solid depends on the height distribution of the asperities but only weakly on the pressure~\cite{zhuravlev2007,gw1966}. Furthermore, for the PAAm suspension the effective suspension friction coefficient is a monotonically increasing function of the shear rate.

\newpage

Gelatin particles suspensions always have a significant effective friction coefficient in the limit of zero shear rate. The suspension friction coefficient also seems to be of the same order as the microscopic friction coefficient. Upon increasing the shear rate, the suspension friction coefficient however initially decreases before entering the more commonly observed rate dependent regime; the larger $\mu_m$, the stronger the decrease. Initially, the decrease seems logarithmic, yet there is always a pronounced minimum in $\mu_m(\dot{\gamma})$. Note that the location of the minimum is at constant $\dot{\gamma}$ for each material, rather than at a constant inertial number~\cite{dacruz2005} $I = \dot{\gamma}d\sqrt{\rho/P^p}$ or viscous number~\cite{boyer2011} $J = \frac{\eta_f \dot{\gamma}}{P^p}$, where $\rho$ is the particle density and $\eta_f$ the viscosity of the suspending fluid. To highlight this fact, we plot $\mu$ as a function of $\dot{\gamma}/P^p$ in the inset of Fig.~\ref{fig:MuRate}c. The collapse of the data is certainly not as good as in the main panel, especially in the slow flow limit. The shear rate at which the minimum occurs thus seems to change little with $P^p(\dot{\gamma})$. At higher gelatin concentration, the particles also change in stiffness by a factor 10 as documented in Sec.~\ref{sec:mnm}, whereas the location of the minimum does not appear to systematically change in panel Fig.~\ref{fig:MuRate}b-d. The role of particle stiffness is perhaps not always crucial in slow flows~\cite{Coulomb2017}, but the absence of good rescaling with either $P^p$ or $E$ suggests that another, perhaps contact-based, time scale is causing the instability.

\begin{figure}[!t]
\centering
\includegraphics[width=0.9\linewidth]{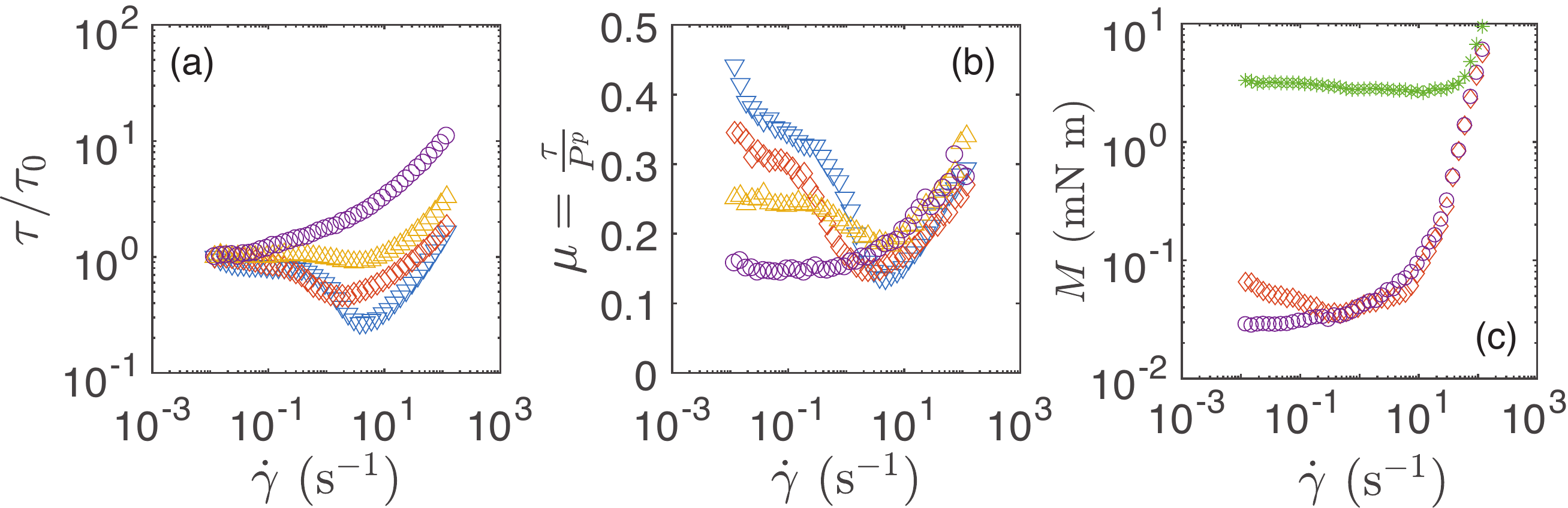}
\caption{\label{fig:comparison} (a) Shear stress $\tau$ normalized with the yield stress $\tau_0$ as function of shear rate $\dot{\gamma}$ for suspensions at fixed volume. We estimate $\tau_0$ as $\tau$ at the lowest $\dot{\gamma}$ considered here. The PAAm suspension ($\circ$, $P^p = 0.24$~kPa) is fitted well by the HB model (see Fig.~\ref{fig:flowcurves}a), while gelatin suspensions (with 15 ($\triangledown$, $P^p = 0.79$~kPa), 10 ($\diamond$, $P^p = 0.87$~kPa), or 5 wt\% gelatin ($\triangle$, $P^p = 0.30$~kPa)) display a flow instability. (b) For the same data as in (a), a comparison of $\mu(\dot{\gamma})$ from PAAm and the three gelatin suspension types. (c) Torque $M$ as a function of shear rate $\dot{\gamma}$ for gravitational suspensions of 10 wt\% gelatin ($\diamond$), PAAm ($\circ$) and glass beads ($*$). All measurements performed in our Couette cell but now filled to a height of $\approx 3/4h$, i.e. there is no pressure on the lid and the particles are jammed by hydrostatic pressure $P_g$ only. Due to the larger density of the glass beads, their yield stress is also larger.}
\end{figure}

\subsection{Comparison of flow curves}
We can go a step further and directly compare the flow curves of different hydrogel suspensions in one figure. We would like to stress that while we change the hydrogel chemistry, all other particle and suspension characteristics such as hardness, size and polydispersity, system volume, boundary conditions et cetera are the same between a PAAm particle suspension and a gelatin particle suspension composed with particles made from a 10\% gelatin solution. We first compare the shear stress behavior in Fig.~\ref{fig:comparison}a; to make a good comparison, we normalize the data on the zero-shear stress value. This allows us to even more directly compare the role of particle hardness: there is no observable trend with the particle modulus in the location of the minimum in the flow curve for the three gelatin-based suspensions, so the flow curves do not seem to be affected by this pressure scale. The depth of the minimum however increases with increasing polymer concentration and thus seems to depend on $\mu_m$.

When we compare the suspension friction coefficients in Fig.~\ref{fig:comparison}b, we see that the four different suspensions behave similar in the high flow rate regime; at low flow rates, the observed minimum in the flow curve for gelatin coincides with the effective friction coefficient for the PAAm suspension. Indeed, recent numerical work suggests that $\mu(\dot{\gamma})$ should depend on $\mu_m$~\cite{trulsson2017}, contradicting earlier results that suggest that $\mu$ is a universal function~\cite{gallier2014}. Here, we observe that at sufficiently high $\dot{\gamma}$, the gelatin data is clearly similar to that of the PAAm data. Note that the exponent $n$ seems to be much less than $1$, contrary to expectations for inertial or viscously damped granular flows, but in agreement with dense emulsion flows~\cite{dinkgreve2015}.

\subsection{Volume control versus pressure control}\label{subsec:vvspcontrol}
In the experiments discussed above, we exclusively focused on volume controlled experiments, in which the volume fraction is fixed and the granular pressure and shear stress depend on the shear rate. This is potentially problematic, as $\phi$ is considered the slaved variable in most flow modeling efforts~\cite{boyer2011,henann2013}. However, the flow behavior we have observed is not limited to controlled volume contexts. We can drive our granular emulsions also without the presence of the confining lid and observe the same qualitative behavior. Without lid, the confining pressure scale is then the hydrostatic pressure generated by the density mismatch of the particles and the water. This constant pressure environment also allows us to compare our granular emulsions with a suspension of glass beads in water. The results are shown in Fig.~\ref{fig:comparison}c. Clearly, in the pressure controlled environment of the open Couette cell, we observe that \textit{i} the PAAm suspension has a monotonically increasing flow curve, \textit{ii} the gelatin suspension has a minimum in the flow curve. \textit{iii} the glass bead suspension shows a modest shear weakening behavior conform to other work~\citep{dijksman2011}. Note that the nonmonotonicity of gelatin suspensions even reproduces in unconfined split bottom geometry~\cite{2010_sm_dijksman} driven flows (not shown). We thus conclude that volume control in our experiments did not significantly change the coupling between microscopic contact mechanics and macroscopic flow phenomenology, again suggesting some other contact based time scale is determining the instability and/or the difference between the PAAm and gelatin suspensions. 

\begin{figure}[!tb]
\centering
\includegraphics[width=0.9\linewidth]{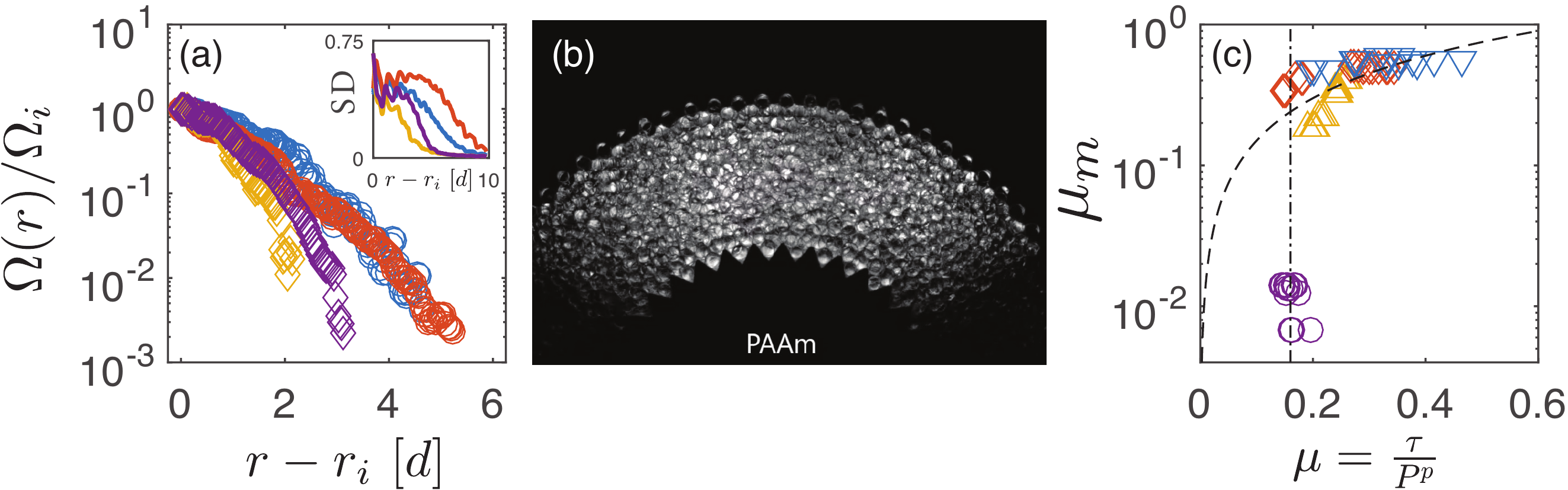}
\caption{\label{fig:flowprofiles} (a) Normalized angular velocity $\Omega(r)/\Omega_i$ as a function of the distance from the inner cylinder for gelatin ($\diamond$) and PAAm ($\circ$) suspensions at similar $P^p \approx$ 0.2~kPa. Yellow and blue datapoints represent the suspension at $\dot{\gamma} = 4.8 \times 10^{-2}$ s$^{-1}$, while purple and red datapoints are at $\dot{\gamma} = 1.2$ s$^{-1}$. Adapted from Ref.~\cite{workamp2017}. Inset: normalized standard deviation $SD$ of the pixel intensity time series, as a function of the distance from the inner cylinder. Same color coding as in the main panel. (b) Still image from supplementary video showing the difference in particle flow fluctuations in PAAm and 10\% gelatin suspensions (Multimedia View). (c) Material friction coefficient $\mu_m$ as a function of the effective friction coefficient of the suspension $\mu$ for all materials, sliding velocities and volume fractions. The dashed line represents $\mu_m=\frac{3}{2}\mu$, dash-dotted line denotes the critical value $\mu_0$. Same symbols as in Fig.~\ref{fig:comparison}.}
\end{figure}

\section{Flow behavior}\label{sec:flow}
\subsection{Flow profiles} 
The Herschel-Bulkley behavior observed for the PAAm suspension, and the flow instability observed for gelatin suggest that our granular emulsions will show shear banding~\cite{schall2010, bonn2017review}. To determine the flow profiles in the gap of our Couette cell, we perform particle image velocimetry (PIV), using a method described in more detail elsewhere~\cite{workamp2017}. In short: imaging the flowing suspension in transmission provides sufficient contrast to elucidate local velocities using standard PIV methods. In Fig.~\ref{fig:flowprofiles}a we plot the angular velocity $\Omega(r)$ normalized with the angular velocity of the inner cylinder $\Omega_i$, for suspensions of PAAm and 10\% gelatin at similar pressure ($P^p \approx$ 0.2~kPa), for two different driving rates: $\dot{\gamma} = 4.8 \times 10^{-2}$ s$^{-1}$ and $\dot{\gamma} = 1.2~$s$^{-1}$, both below $\dot{\gamma}$ of the minimum. A more extensive dataset can be found in Ref.~\cite{workamp2017}. For the PAAm suspensions, the shear bands are relatively wide and insensitive to the driving rate, whereas for gelatin the shear bands are narrow and slightly rate-dependent. However, for both materials, $\Omega(r)$ decays to less than 10\% of $\Omega_i$ within a few particle diameters $d$.

\subsection{Flow Fluctuations} 
Even though the decay of the velocity profiles shows that flow ceases entirely beyond a couple of particle diameters from the rotating cylinder, we observe that the suspended particles still fluctuate in their position even in the static zone. That PAAm and gelatin suspensions display different fluctuations can be observed visually when running the experiment. Particles outside of the shear band are clearly much more ``agitated'' in a PAAm suspension compared to the gelatin case. These velocity fluctuations are a crucial element in dispersion based flow modeling~\cite{jop2012,kamrin2017}. Measuring the actually relevant particle-level velocity fluctuations is not possible in our experiment, but we can qualitatively measure the extent of such fluctuations. We estimate particle position fluctuations by calculating the standard deviation $SD$ of the time series of the intensity fluctuations in the forward scattered light passing through the suspension. Values are averaged over the azimuthal direction and normalized with the mean intensity of the image. $SD$ signifies both flow and uncorrelated particle motion, and is plotted in the inset of Fig.~\ref{fig:flowprofiles}a, for the same experiments represented in the main panel. From this analysis, two observations stand out: it is clear that particle fluctuations extend the entire gap ($r-r_i\approx 10d$) for the PAAm suspension, even though the flow is localized to a shear band of only a few particle diameters. By contrast, in the case of gelatin, $SD$ decays to zero within 1 or 2 particle diameters away from the shear band. Second, the extent of particle motion fluctuations for the PAAm suspensions are rate dependent, whereas the normalized flow profiles are not. These observations can be qualitatively assessed with multimedia video~\ref{fig:flowprofiles}b. We conclude that the microscopic frictional interaction mechanisms also significantly affect the local flow behavior: friction enhances shear banding and suppresses particle-level fluctuations.\\ 

\subsection{Microscopic interpretation}
The measured flow profiles allow us to estimate the relative velocities of the particles in the suspension. Since the microscopic friction is weakly rate dependent for the gelatin particles we can attempt to see how $\mu_m(v)$ and $\mu(\dot{\gamma})$ are connected. Since $\Omega(r)$ decays to less than 10\% of $\Omega_i$ within a few particle diameters $d$, we estimate the particle relative sliding velocities as $\Omega_i r_i$. The sliding velocities in our friction measurements are then corresponding to values of $\dot{\gamma}$ just below the observed minima in $\mu(\dot{\gamma})$. We can therefore speculate that a microscopic timescale in $\mu_m(v)$ plays a role in the observed flow instability. The observed irrelevance of pressure versus volume control in Sec.~\ref{subsec:vvspcontrol} points towards a more microscopic underpinning of the observed minimum. Our data is however inconclusive: the strongest instability is observed for the 15\% gelatin particle, which shows the least amount of rate dependence in $\mu_m(v)$. Due to experimental limitations, we cannot extend the range of $\mu_m(v)$ to higher $v$, covering the entire $\dot{\gamma}$ range. We can nevertheless directly compare our $\mu_m(v)$ and $\mu(\dot{\gamma})$ by plotting $\mu_m(v)$ for each $\mu(\dot{\gamma})$ with a similar sliding velocity. We plot $\mu_m$ as a function of $\mu$ in Fig.~\ref{fig:flowprofiles}c, for all materials and $P^p$. The dashed line serves as a reference to indicate what a linear relation between the two variables would look like on this log-log scaling; specifically, it represents $\mu_m=\frac{3}{2}\mu$. For the gelatin, all data points lie close to this line: the collapse is certainly not perfect, but the deviations in $\mu$ are all smaller than 0.15. This is important, as for the PAAm suspensions, at much lower $\mu_m = 0.01$, $\mu$ differs distinctly from $\mu_m$, by as much as $0.15$. This deviation suggests that a different dissipation mechanism must contribute to the shear resistance of the PAAm suspension, that sets a minimum $\mu$, which we call $\mu_0$. We find $\mu_0 \approx 0.16$. In simulations of slow flows of frictionless suspensions~\cite{chialvo2012,trulsson2017} and frictionless dry granular materials~\cite{peyneau2008}, values of approximately 0.1 are found, suggesting a bigger contribution of fabric and force anisotropy~\cite{rothenburg1989,majmudar2005,peyneau2008,peyneau2008b,azema2014,azema2015} in our experiments. Unexpectedly, for frictional particles like the gelatin particles used here, the correction due to anisotropy/geometry seems to disappear, and the suspension friction coefficient is set exclusively by the material friction coefficient. Knowing that even in the PAAm suspension a finite $\mu_0$ is observed, even though tangential contact force components are absent makes it all the more surprising that, approximately, $\mu_m=\frac{3}{2}\mu$ for the gelatin suspensions: the contributions of friction and geometric effects to the shear stress do not seem to be simply additive; it seems that $\mu = \mu_m + C(\mu_m)$ in which constant $C(\mu_m) \sim \mu_0$ for $\mu_m \ll 0.1$, but $C(\mu_m) \rightarrow 0$ for $\mu_m > 0.1$  Adding to the confusion, numerical simulations of comparable systems have found contradicting relationships between $\mu_m$ and $\mu$: see Refs.~\cite{chialvo2012,gallier2014,trulsson2017}. Finding how anisotropy emerges from grain-scale friction, velocity and perhaps other microscopic contact and force correlations hence seems to be an important next step to understand suspension rheology.

What is the microscopic source of the instability? We would like to note that the instability observed in Fig.~\ref{fig:comparison}c for the glass beads and gelatin suspension is of different character. The glass bead suspension has a logarithmic negative rate dependence, that beyond a certain flow rate gets overtaken by inertial dynamics. The source of this rate dependence is perhaps related to self-weakening due to mechanical agitations present in the material~\cite{wortel2016} that propagate fast enough due to the hardness of the particles and the limited damping of the low viscosity solvent (water). In contrast, the gelatin suspension has a broad and deep minimum in the flow curve. The time scale responsible for this minimum is not clear. One option is that it is related to a hydrodynamic particle contact effect. Due to the composition of the hydrogel particles used in this study, probing the role of the fluid viscosity was not possible, yet this remains a promising avenue for future work. \\


\section{Conclusions} 
We probe the role of microscopic friction in slow dispersed media flows by synthesizing soft particles that allow us to perform experiments above the random close packing limit. By measuring both shear and confining stresses during flow, we find that friction plays an outsize role in all aspects of the flow: rheology, flow profiles and particle-level fluctuations of such suspensions are significantly affected by the microscopic friction coefficient. In the ``emulsion'' limit, where the material friction friction $\mu_m$ is smaller than a critical value $\mu_0$, the macroscopic friction $\mu$ remains finite. This suggests that dissipation in dispersed media can emerge from non-frictional, perhaps geometric sources or velocity fluctuations. Upon increasing the material friction coefficient $\mu_m > \mu_0$, we find that the flow behavior of the granular emulsion becomes unstable, while the effective friction coefficient of the suspension approaches that of the microscopic value; we find that the suspension friction coefficient $\mu$ is set by $\frac{2}{3}\mu_m$. Our results show that the ``granular emulsion'' phase yields a wide range of different, unexpected and potentially useful flow behaviors. The observations provide new benchmarks for modeling approaches and could serve as input to get more insight in the microscopic underpinning of fluidity and anisotropy based modeling of dispersed media.  

\begin{acknowledgments}
We thank Pieter de Visser, Raisa Rudge and Jonathan Bar\'{e}s for their help and useful discussions concerning the tribology measurements, and Sepideh Alaie for her help with making the particles and measuring the flow profiles.
\end{acknowledgments}
%
\end{document}